\documentclass[aps,pra,floatfix,superscriptaddress,showpacs]{revtex4}

\usepackage{graphicx}

\begin{document}

\title{A microwave transducer for a  nano mechnical resonator.}

\author{G.J.Milburn, C.A.Holmes, L.M. Kettle} 
\affiliation{School of Physical Sciences\\
	     The University of Queensland, St Lucia, Australia}
\author{H.S. Goan}
\affiliation{Department of Physics,
National Taiwan University, Taipei 106, Taiwan, ROC}

\begin{abstract}
We give a quantum master equation description of the measurement scheme based on a coplanar  microwave cavity capacitively coupled to  nano mechanical resonator. The system exhibits a rich bifurcation structure that is analogous to sub/second harmonic generation in nonlinear optics. We show how it may be configured as a bifurcation amplifier transducer for weak force detection. 
\end{abstract}

\maketitle

\setcounter{section}{0}

\section{Introduction}
A Quantum Electromechanical System (QEMS) is a sub micron fabricated mechanical system, with action of the order of Planck's constant and incorporating electronic transducers operating at the quantum limit\cite{QEMS_PT}. Such devices may enable a force microscopy sensitive enough to detect the magnetic moment of a single spin, or the deformation forces on a single macromolecule with applications to information processing and biomolecular technology.  

Current technology now enables the fabrication of QEMS with resonant frequencies approaching 1 GHz\cite{Roukes1} and at milli Kelvin temperatures such an oscillator can be close to its vibrational ground state. La Haye et al.\cite{LaHaye} described a measurement of a QEMS displacement with a sensitivity approaching the Heisenberg limit. The ability to fabricate a QEMS at GHz frequencies motivates the discussion presented below in which we describe a QEMS transducer based on super-conducting microwave circuits.

Coplanar super-conducting  wave guides are currently of great interest for quantum information processing in solid states systems\cite{Walraff}.   Our objective here is to capacitively couple the mechanical oscillator of a QEMS directly to the cavity field. We show that  if the mechanical frequency is twice that of the circuit resonance, the system is completely analogous to the nonlinear optical model of sub/second harmonic generation.

What is a quantum electromechanical system? At the Solvay Conference in 1927, Einstein challenged Bohr to explain the developing quantum theory in realistic terms. Of this debate, Bohr said,\cite{Bohr}, ``the discussions...centered on the question 
of  whether the quantum-mechanical  description exhausted the possibilities of  accounting for the observed phenomenon... Einstein maintained the analysis could  be carried further...by bringing into 
consideration the detailed balance of  energy and momentum in individual  processes."  As a result of the discussion, Bohr proposed various mechanical `thought experiments' which would clarify his position. Bohr's imaginary apparatus could well be the first proposals for quantum electromechanical systems. 
%
However to describe such macroscopic devices in quantum theoretical terms would be a gross extravagance. In one of Bohr's schemes a screen with a single slit is mounted on springs: a mechanical resonator. The idea is to detect the energy transferred to the screen when a diffracted quantum particle is found at a distant detector. Thermal fluctuations in the motion of the oscillator-screen will mask such transfers, so it would be provident to cool the system close to the ground state. This will require $\hbar\nu>k_BT$ where $\nu$ is the mechanical resonance frequency. Under the most optimistic circumstances, the temperature is bounded below by milli Kelvin which would then require a mechanical resonance in the gigahertz regime: a microwave frequency! Clearly this is not an everyday regime for a mechanical oscillator. 

In order to see how to make such a high frequency resonator we note that the mechanical resonant frequency will scale as the inverse of the length scale, $\nu\propto l^{-1}$. For example, a doubly clamped bar of SiC with dimensions $w\times l\times t=100\times 3\times 0.1\  \mu$m has  $\nu=120$kHZ while one with dimensions of $0.1\times0.01\times0.01\  \mu$m has $\nu=12$Ghz\cite{Roukes2}. We see that to build a mechanical resonator in which quantum effects begin to dominate we must look to fabricating nano-mechanical systems. 

With a device that small how does one build a transducer for the motion of the resonator? Optical methods are widely used to transduce the motion of MEMS devices. This will not work well at the nanometer scale as the objects we seek to image are at or below the diffraction limit. No doubt there are some near-field tricks that might be used to fix this, however much recent work has concentrated  on using single-electronics to detect the motion. The idea is to enable the nano-mechanical resonator to modulate a circuit capacitance and thereby modulate the gate voltage on a single electron transistor\cite{LaHaye} .   In this paper we propose a different kind of transducer based on super-conducting co planar microwave cavities. This approach is strongly indicated by the gigahertz frequency range of a quantum nano-mechanical system.

\section{A microwave nano-mechanical motion transducer.}
The Hamiltonian for the electronic circuit in Fig.{\ref{NEMS-cap} is 
\begin{equation}
H=H_{osc}+\frac{\hat{P}^2}{2L_T}+\frac{\hat{Q}^2}{2(C_T+C_o(\hat{x}))}-e(t)\hat{Q}
\end{equation}
where the canonical variables , $\hat{P},\hat{Q}$, are the current in the inductor and charge on the capacitor respectively\cite{Louisell}, $L_T,C_T$ are the inductance and capacitance of the equivalent tank circuit, $C_o(\hat{x})$ is the capacitance due to the coupling to the mechanical oscillator with $\hat{x}$ the coordinate of small oscillations of the mechanical oscillator around its equilibrium position and $e(t)$ is a classical ac driving voltage to the tank circuit. The free Hamiltonian of the mechanical oscillator is $H_{osc}$. 
\begin{figure}[htbp]
\centering
\includegraphics[scale=0.5]{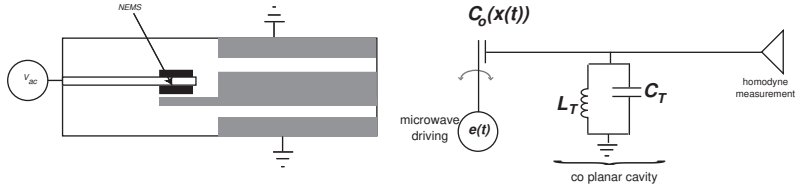}
\caption{\label{NEMS-cap} A direct capacitive coupling between a nano-mechanical resonator and a co planar microwave cavity. The mechanical resonator is doubly clamped between the gap indicated by the black region. Also shown is an equivalent circuit model. It forms a time dependent capacitive coupling between the input transmission line driven by a microwave source voltage.   }
\end{figure}
This model is very similar to that proposed by Blencowe and Wybourne\cite{blencowe_squeeze} to discuss squeezing due to parametric driving,  however here we treat the circuit as a distinct quantum mechanical oscillator.

We assume that we may expand $C_o(\hat{x}) $ as 
$
C_o(\hat{x}) =C_o+\mu\hat{x}(t)
$
with $\mu=\left .\frac{dC(x))}{dx}\right|_x=0$. On the assumption that $C_\Sigma=C_T+C_o >>\mu\langle\hat{x}\rangle$ we further expand the capacitive energy term in the Hamiltonian to first order in in $\hat{x}$ to obtain
\begin{equation}
H=\frac{\hat{p}^2}{2M}+\frac{M\omega_m^2}{2}\hat{x}^2+\frac{\hat{P}^2}{2L_T}+\frac{\hat{Q}^2}{2C_\Sigma}+ \frac{\mu}{2 C_\Sigma^2} \hat{Q}^2\hat{x}-e(t)\hat{Q} - f(t)\hat{x}
\end{equation}
and we have now included the free hamiltonian of the mechanical oscillator with mass $M$ and frequency $\omega_m$, together with a mechanical forcing term $f(t)$.  The LC circuit oscillation resonant frequency is $\omega_e=(L_TC_\Sigma)^{-1/2}$. We will assume that the electronic driving, $e(t)$, and the mechanical driving, $f(t)$, are harmonic with $e(t)=2\hbar\epsilon_e\cos(\omega_Dt),\ \ f(t)=2\hbar f_m\cos(\Omega t)$. 

The key feature to note in this Hamiltonian is the nature of the coupling between the electronic and mechanical oscillators: it is linear in the oscillator coordinate but {\em quadratic} in the electronic coordinate. This is a parametric coupling analogous to that found in sub-second harmonic generation in nonlinear optics\cite{Walls_Milb}. Such systems have a rich dynamical structure including stable and unstable fixed points and limit cycles, and the operation of this system as a position transducer requires some care. 

We now define bosonic lowering operators for the circuit and mechanical oscillator, respectively
$b  =  \frac{1}{\sqrt{2\hbar\omega_e L_T}}(\omega_eL_T \hat{Q}+i\hat{P})\ \ \ \ a  =  \frac{1}{\sqrt{2\hbar M\omega_m}}(M\omega_m \hat{q}+i\hat{p})$.
The Hamiltonian now takes the form
\begin{eqnarray}
H& = &\hbar\omega_eb^\dagger b+\hbar\omega_ma^\dagger a+\hbar\chi(b+b^\dagger)^2(a+a^\dagger) \nonumber \\
& & \mbox{}-\hbar\epsilon_1(b+b^\dagger)(e^{i\omega_D t}+ e^{-i\omega_D t}) - \hbar \epsilon_2(a+a^\dagger)(e^{-i\Omega t}+e^{i\Omega t})
\end{eqnarray}
where $\chi = \left(\mu \omega_e / 4 C_\Sigma \right) \sqrt{\hbar / (2 \omega_m M)} $, $\epsilon_1 = \epsilon_e \sqrt{\hbar / (2 \omega_e L_T)}$, and  $\epsilon_2 = f_m \sqrt{ \hbar / (2 \omega_m M)}$.

We assume the device has been engineered so that the resonance condition $2\omega_e=\omega_m$ applies.
Moving to an interaction picture and neglecting counter rotating terms in the circuit-oscillator interaction term, we obtain the interaction picture Hamiltonian,
\begin{equation}
H_I(t)=\hbar\chi\left (a^\dagger  b^2 + a(b^\dagger)^2\right )-\hbar\epsilon_1(be^{i\Delta t}+b^\dagger e^{-i\Delta t}) - \hbar \epsilon_2 (ae^{i\delta t}+a^\dagger e^{-i\delta t}) \label{eq:1}
\end{equation}
where $\Delta=\omega_D-\omega_e$ and $\delta=\Omega - \omega_m$.  

In addition to the explicit electronic and mechanical degrees of freedom included in the Hamiltonian we need to model dissipation and associated noise sources.  We will use the quantum optics master equation for both the electronic and mechanical dissipation. This is expected to be adequate so long as both oscillators are very under-damped.   The resulting master equation (in the interaction picture) for the total state of electronic plus mechanical oscillators is then\cite{Walls_Milb}
 \begin{eqnarray}
\frac{d\rho}{dt} & = & -\frac{i}{\hbar}[H_I(t),\rho]+\gamma_e(\bar{n}_e+1){\cal D}[b]\rho+\gamma_e\bar{n}_e{\cal D}[b^\dagger]\nonumber \\
  & & \mbox{}+\gamma_m(\bar{n}_m+1){\cal D}[a]\rho+\gamma_m\bar{n}_m{\cal D}[a^\dagger]
  \label{ME}
\end{eqnarray}
where $\gamma_e,\ \ \gamma_m$ are the amplitude damping rates of the electronic and mechanical oscillators respectively, $\bar{n}_e,\ \ \bar{n}_m$ are the mean bosonic occupation numbers for the electronic and mechanical baths ( not necessarily at the same temperature).  The super operator ${\cal D}$ is defined by
\begin{equation}
{\cal D}[A]\rho=A\rho A^\dagger -\frac{1}{2}\left (A^\dagger A \rho +\rho A^\dagger A\right )
\end{equation}
The master equation, Eq.(\ref{ME}), has been previously used to describe sub-second harmonic generation for two quantised fields interacting in a medium with a significant second order optical nonlinearity\cite{Walls_Milb,DMcNW}}. The first step to understanding the behavour is to consider the semiclassical equations of motion. These exhibit a rich behaviour including fixed points, and limit cycles.  

\begin{figure}[htbp]
\centering
\includegraphics[scale=0.4]{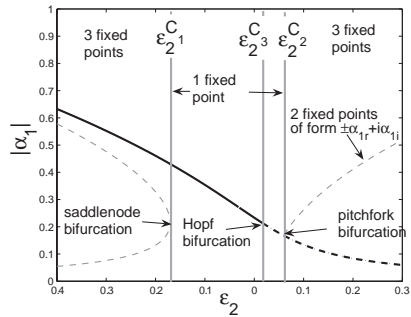}
\caption{\label{stability} A stability diagram for the coupled system in terms of the value of the dimensionless electric field magnitude at the fixed points as a function of the mechanical force on the nano-mechanical resonator. }
\end{figure}

The classical non-linear equations of motion (on resonance $\Delta=0,\delta=0$) for second harmonic generation are:
\begin{eqnarray}
\dot{\alpha_1} & = &  2 \imath \chi \alpha_1^* \alpha_2 - \imath \epsilon_1 - \frac{\gamma_1}{2} \alpha_1, \\
\dot{\alpha_2} & = &  \imath \chi \alpha_1^2 - \imath \epsilon_2- \frac{\gamma_2}{2} \alpha_2,
\label{e01}
\end{eqnarray}
together with their complex conjugate counterparts. Here $\alpha_1$ and $\alpha_2$ are proportional to the amplitudes of the circuit and cantilever respectively. This system of equations has a rich dynamical structure in terms of fixed points and limit cycles. Figure \ref{stability} is a partial summary of the conditions. Here we wish to focus on the Hopf bifurcation that takes the fixed point to a limit cycle transition as $\epsilon_2$ is increased above a critical value: $\epsilon^{cr}_2$, marked as `hopf bifurcation' in the figure. 

\section{A bifurcation amplifier for weak force detection.}
Returning to one of the primary applications for a nanomechanical system, the detection of a weak force, we imagine a scenario in which the weak force we wish to detect is fixed in size but varies in sign. That is to say, we wish to know if the force is attractive or repulsive. As an example, we could imagine we wish to detect whether a given quantum dot is positively or negatively charged with respect to some potential background.  The size of the charge could be fixed at $e$ and we only wish to determine the sign. This is the situation that occurs for charge qubit realisations\cite{Hollenberg}. A more important example is to detect the direction of a single spin. In the case of an electron, the magnitude of the force is fixed given a fixed magnetic field gradient at the spin, but the sign depends on the direction of the spin with respect to a static magnetic field.  If the field gradient is caused by a nano-magnet on the nano-mechanical system, we really are only interested to know if the  force on the resonator is positive or negative. 

We can configure the coupled circuit/nano-mechanical system to achieve this objective using an approach of Siddiqi et al.\cite{siddiqi}: the bifurcation amplifier. This concept has recently been used to readout a super conducting qubit in a nonlinear circuit where the nonlinearity arise form Josephson tunneling. Here this nonlinearity is replaced by the nonlinear coupling of the nanomechanical resonator and the circuit. Indeed, this approach is quite general: incorporating a nano-mechanical resonator into a super conducting wave guide is a viable alternative approach to using the nonlinear response of the Josephson junction.

\begin{figure}[htbp]
\centering
\includegraphics[scale=0.5]{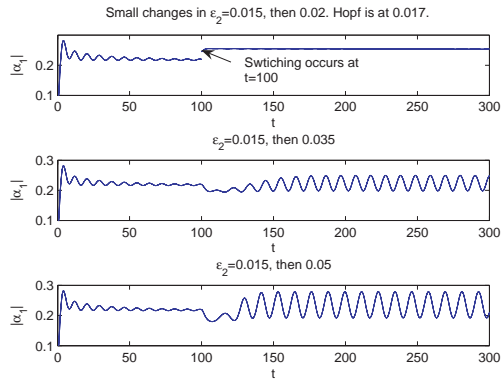}
\caption{\label{switching} The magnitude of the dimensionless electric field in the cavity versus time, with the force on the nano-mechanical system changed by a discrete amount at $t=100$. BY this time the system has settled onto a fixed point.  In (a) the mechanical force is reduced, in (b) and (c) the mechanical force is increased.  Time units are chosen so that $?\chi=1$. }
\end{figure}

Suppose we choose the operating conditions so that the system has settled onto a fixed point just before the Hopf bifurcation. If the mechanical force parameter $\epsilon_1$ is {\em increased} by a discrete step, the system will switch to the limit cycle if the change is big enough. On the other hand if the force is {\em reduced} by a discrete step, the system will simply adjust to a new  stable fixed point. Thus the sign of the change in the force is signaled by a self sustained oscillation of the electric field in the co planar cavity, a signal that can be transduced using heterodyne detection.   A number of examples are shown in figure \ref{switching}. 

Note that this approach is practical only if the change in the force is limited to a single discrete step. Of course in reality the magnitude  of the step will fluctuate due to noise. In order  not to turn the construction into a noise amplifier we need to ensure that any physical reasonable level of noise will not switch the system beyond the Hopf bifurcation by itself. This will determine how small a step we can detect and depends on how close to the Hopf bifurcation we can operate in the presence of noise.  How small a force can this approach  resolve? A detailed answer to this question requires that we include the stochastic terms in the equations of motion.


\begin{thebibliography}{00}
\bibitem{QEMS_PT}K.C.Schwab, M.L.Roukes, Putting Mechanics into Quantum Mechanics, 
Physics Today, JULY, (2005).
\bibitem{Roukes1}X.M.H.Huang, C. A. Zorman, M. Mehregany, M.L. Roukes, Nature,  {\bf 421} ,496, (2003).
\bibitem{Roukes2}M.L.Roukes, Technical Digest of the 2000 Solid-State Sensor and Actuator Workshop, Hilton Head Isl., SC, 6/4-8/2000 (ISBN 0-9640024-3-4).
\bibitem{LaHaye}M. D. LaHaye et al., Science 304, 74 (2004).
\bibitem{Walraff}Walraff et al., Nature {\bf 431}, 162,  (2004).
\bibitem{PP}D.N. Jamieson, C. Yang, T. Hopf, S.M. Hearne, C.I. Pakes, S. Prawer, M. Mitic, E. Gauja, S.E. Andresen, F.E. Hudson, A.S. Dzurak, and R.G. Clark,  Applied Physics Letters, {\bf 86}, 202101 (2005). 
\bibitem{Jacobs}Asa Hopkins, Kurt Jacobs, Salman Habib, and Keith Schwab, Phys. Rev. B  {\bf 68}, 235328 (2003).
\bibitem{Bohr}N.Bohr, `Discussions with Einstein on epistemological problems in atomic physics', in {\em Albert Einstein Philosopher-Scientist}, 1: 201 ed. P. Schilpp (New York: Harper \& brothers, 1959). 
\bibitem{Louisell}W.H.Louisell, {\em Quantum statistical properties of radiation}, (Wiley-Interscience, 1990)
\bibitem{blencowe_squeeze}M. P. Blencowe, M. N. Wybourne, Physica B, {\bf 280},555 (2000). 
\bibitem{Walls_Milb}D.F.Walls and G.J.Milburn, {\em Quantum Optics}, (Springer, Berlin 1994). 
\bibitem{DMcNW}P.D. Drummond, K.J. McNeil, D.F. Walls, Optica Acta, {\bf 28}, 211 (1981). 
\bibitem{Hollenberg}L. C. L. Hollenberg, A. S. Dzurak, C. Wellard, A. R. Hamilton, D. J. Reilly, G. J. Milburn, and R. G. Clark, Phys. Rev. B 69, 113301 (2004)  
\bibitem{siddiqi}I. Siddiqi, R. Vijay, M. Metcalfe, E. Boaknin, L. Frunzio, R. J. Schoelkopf, and M. H. Devoret, Phys. Rev. B 73, 054510 (2006).





\end{thebibliography}
\end{document}